\documentclass
      [reprint,amsmath,amssymb,superscriptaddress,groupedaddress]{revtex4-1}

\usepackage{graphicx} % Include figure files
\usepackage{braket}
\usepackage{amsmath}
\usepackage{amssymb}

\input epsf
\input rotate

\def\bea{\begin{eqnarray}}
\def\eea{\end{eqnarray}}
\def\ben{\begin{equation}}
\def\een{\end{equation}}
\def\benu{\begin{enumerate}}
\def\enu{\end{enumerate}}

\def\n{n}
\def\lsim {\ifmmode {\buildrel<\over\sim}}

\def\sss{\scriptscriptstyle\rm}

\def\1var{(\bx_1...\bx\N)}

\def\br{{\bf r}}

\def\b1{{\bf 1}}
\def\bx{{x}}

\def\xc{_{\sss XC}}

\def\N{_{\sss N}}
\def\H{_{\sss H}}

\def\ext{_{\rm ext}}

\def\sph_int{ {\int d^3 r}}

\def\infintd3r{ \int_{-\infty}^\infty d^3r\,}
\def\intd3r{ \int d^3r\,}

\def\laplace1d{\frac{d^2}{dx^2}}
\def\plaplace1d{\frac{d^2}{d{x'}^2}}

\def\padr2{\frac{\partial^2}{\partial r^2}}

\def\N{{\cal N}}

\def\b{{\beta}}

\begin{document}

\title{Accurate Reference Data for the Non-Additive Non-Interacting Kinetic Energy in Covalent Bonds}
\author{Jonathan Nafziger} 
\affiliation{Department of Chemistry, Purdue University, 560 Oval Dr., West Lafayette IN 47907, USA}
\author{Kaili Jiang} 
\affiliation{Department of Physics, Purdue University, 525 Northwestern Ave., West Lafayette, IN 47907, USA}
\author{Adam Wasserman}
\email{awasser@purdue.edu}
\affiliation{Department of Chemistry, Purdue University, 560 Oval Dr., West Lafayette IN 47907, USA}
\affiliation{Department of Physics, Purdue University, 525 Northwestern Ave., West Lafayette, IN 47907, USA}

\begin{abstract}
The non-additive non-interacting kinetic energy is calculated exactly for fragments of H$_2$, Li$_2$, Be$_2$, C$_2$, N$_2$, F$_2$, and Na$_2$ within partition density-functional theory.  The resulting fragments are uniquely determined and their sum reproduces the Kohn-Sham molecular density of the corresponding XC functional.  We compare the use of fractional orbital occupation to the usual PDFT ensemble method for treating the fragment energies and densities.  We also compare Thomas-Fermi and von Weiz{\"a}cker approximate kinetic energy functionals to the numerically exact solution and find significant regions where the von Weiz{\"a}cker solution is nearly exact.
\end{abstract}

\maketitle

\section{Introduction}
The non-additive non-interacting kinetic energy (NAKE) and its functional derivatives are critical functionals for density-based fragment methods.  These methods, such as subsystem density-functional theory\cite{JN2014,KSGP2015} (subsystem-DFT) and partition density-functional theory\cite{CW2007,EBCW2010,NW2014} (PDFT) are based around a real space partitioning of the total density into fragments such that,
\ben
n(\br) = \sum_\alpha n_\alpha(\br),
\een   
where the $\alpha$ index runs over all the fragments.  The NAKE functional is defined as the difference between the kinetic energy of the set of fragment densities and the kinetic energy of the system which has the same density as the sum of the fragments densities, $n_f$,
\ben
T_s^{\rm nad}[\{\n_\alpha\}] = T_s[n_f] - \sum_\alpha T_s[n_\alpha],
\label{eq:Tsnad}
\een 
where the non-interacting kinetic energy, $T_s$, is defined via a constrained search,
\ben
T_s[n] = \min_{\phi\rightarrow n} \bra{\phi}\hat{T}\ket{\phi}
\label{eq:Ts}
\een
Typical approximations to the non-additive quantity of Eq. \ref{eq:Tsnad} are simply based on approximations to the non-interacting kinetic energy functional, which are then plugged into Eq. \ref{eq:Tsnad} to provide an approximation to the non-additive functional.  These approximations have enjoyed success in systems where fragments are not covalently bonded, but it is considered that all known approximations fail for systems where fragments exhibit a strong overlap\cite{WSZ2015}.  Most notably this occurrs when the choice of fragmentation cuts accross covalent bonds.

In recent years, many workers have developed programs to find accurate reference data for the NAKE functional and its derivatives evaluated for specific fragment densities regardless of the strength of the overlap between fragments.  These studies are capable of reproducing Kohn-Sham energies and densities via fragment calculations, with primarily two different goals in mind.  First the corresponding NAKE functional derivatives or non-additive kinetic potentials (NAKPs) are critical ingredients in embedding potentials which allow higher level wavefunction methods to be approximately combined with KS-DFT calculations \cite{HPC2011,GAMMI2010}, thus improving upon the accuracy of KS density functional approximations for some small region of interest within a larger system.

The second goal has been to gain accurate reference data in order to improve density functional approximations for the NAKE and NAKP functionals \cite{DSW2012, FJN+2010, JV2013}.  However, one issue that arises with these calculations within standard subsystem-DFT is that the fragment densities are not uniquely determined when the exact NAKE functional is used.  Thus, for a given system there may be many possible choices of fragment densities whose sum exactly reproduces a KS-DFT calculation for the entire system.  Each of these choices for fragment densities may lead to a different NAKE and also a different NAKP.  On the other hand, with approximate functionals for the NAKE, the self-consistent procedure of subystem DFT defines a unique set of fragment densities for any given system which tends to minimize the fragment overlap\cite{WSZ2015}.  Thus there is a mismatch between exact subsystem-DFT and approximate subsystem-DFT.

Within PDFT the fragments are uniquely defined by requiring that the fragment energies be minimized.  In other words, out of all of the possible fragment densities that solve the subsystem-DFT equations exactly, the PDFT fragments are the ones which have the lowest energy \cite{NW2014}.  This minimization requirement ensures a unique solution and also ends up causing all the fragment embedding potentials to be equal to a single global embedding potential (referred to as the partition potential).  Huang et al. also use a reformulated version of subsystem-DFT with unique fragment densities based around constraining all fragment embedding potentials to be equal \cite{HPC2011}.  For both of these methods the fragment densities are uniquely determined both when the exact and approximate NAKE functionals are used.  Thus these methods provide more convenient reference data when attempting to compare approximate and exact NAKE functionals.

These two methods are nearly identical except for differences in how non-integer spins and charges are handled within fragments.  Huang et al. typically use fractionally occupied orbitals (FOO) when a fragment does not have an integer number of electrons, while PDFT typically uses an ensemble treatment (ENS).  Indeed, there is nothing fundamental to PDFT which prevents the use of FOO fragments nor is there anything in the uniqe formulation of subsystem-DFT which prevents the use of ENS fragments.  Fabiano et al. developed a version of standard subsystem-DFT which uses and ensemble treatment of fragments, allowing for non-integer occupation \cite{FLDS2014}.  The details of the ENS and FOO methods will be presented in the following section and results will be compared in section \ref{results}.

In this paper we provide accurate reference data for both the NAKE functional and its functional derivatives evaluated for the unique PDFT fragment densities for a series of covalently bonded homonuclear diatomics for a variety of separations.

\section{Introduction to PDFT}
Here we will give a brief overview of PDFT.  For a more in depth exposition see reference \citenum{NW2014}.  In PDFT the only choice directly related to the fragmentation of the system is how the nuclear potential of the supersystem is divided into fragment potentials.  In theory, any choice is valid as long as the fragment potentials sum to give the total nuclear potential.  In practice, however, it makes the most sense to assign atoms to various fragments and then the fragment potential is simply the coulomb potential originating from the nuclei of the atoms assigned to each fragment.  In the case of the diatomics studied here each atom is its own fragment.  

Next, $N_\alpha$ electrons are assigned to each fragment.  In the full PDFT formalism this distribution is determined by chemical-potential equalization \cite{TNW2012}, but here we avoid that complication by considering only homonuclear diatomics, where by symmetry an equal number of electrons are assigned to each fragment.  In the covalent bonds we consider here this introduces a slight complication, due to the fact that each fragment will be assigned a number of electrons corresponding to a spin-polarized fragment density.  However, the fragment densities must sum to yield the molecular density which is spin-unpolarized.  The usual PDFT solution in this case is to make the fragment density be represented by a 50-50 ensemble of two spin polarized densities with oppositely polarized densites\cite{NW2015} (ENS).  The other common solution is to have spin-unpolarized fragments with the HOMO orbital only partially occupied (FOO).  We will first introduce the details of the ENS fragments and then the details of the FOO fragments.

\subsection{ENS vs. FOO}
The number of electrons in each fragment ensemble component is $N_{i\alpha\sigma}$, where $i$ is the ensemble component index, $\alpha$ is the fragment index, and $\sigma$ is the spin index.  $N_{i\alpha\sigma}$ will always be an integer number of electrons, but the number of electrons of a given spin for a fragment,
\ben
\label{e:Nalpha}
N_{\alpha\sigma} = \sum_if_{i\alpha}N_{i\alpha\sigma}
\een
will not neccesarily be an integer.  Here, the $f_{i\alpha}$ are the ensemble coefficients, which satisfy the sum rule, $\sum_if_{i\alpha}=1$.  In the covalently bonded cases considered here there will be two equally weighted components in each ensemble, $f_{1\alpha}=f_{2\alpha}=1/2$ and each component will be a spin-flipped version of the other.  In other words $N_{1\alpha\uparrow} = N_{2\alpha\downarrow}$ and $N_{1\alpha\downarrow} = N_{2\alpha\uparrow}$.  The energy of each fragment is the ensemble average of its component energies,
\ben
\label{e:Ealpha}
E_{\alpha} = \sum_if_{i\alpha}E_\alpha[n_{i\alpha\uparrow},n_{i\alpha\downarrow}].
\een
The subscript $\alpha$ on the energy denotes that this is the energy functional corresponding to densities in the external potential $v_\alpha(\br)$ rather than the total external potential.

The alternative option is to consider that the fragments are spin unpolarized and simply allow fractionally occupied orbitals (FOO).  In this case the fragment densities are given by
\ben
n_\alpha(\br) = \sum_j f_{j\alpha}|\phi_{j\alpha}(\br)|^2,
\een
where, $\phi_{j\alpha}$ is the $j^{\rm th}$ Kohn-Sham fragment orbital.  The $f_{j\alpha}$ variable may take on values of 0, 1 or 2 depending on whether the given orbital is unoccupied, partially occupied or fully occupied.  The corresponding fragment energies are given through direct evaluation on the spin-unpolarized fragment density,
\ben
\label{EFOO}
E_{\alpha} = E_\alpha[n_\alpha]
\een
The primary difference to note between ENS and FOO is that in the case of ENS the energy functional is evaluated on spin densities with integer number of electrons in each spin, while in the case of FOO the energy functional is evaluated on spin-unpolarized densities which correspond to non-integer spin densities.  Cohen et al. showed that the exact energy functional will yield the same energy in both cases as long as the total density is the same\cite{CMY2008}, but exchange correlation functional approximations fail to reproduce this behavior and instead wildly overestimate the energy of fractional spins.  This is also the case for approximate NAKE functionals\cite{NW2014}. 

\subsection{Partition Energy and Partition Potential}
In PDFT the sum of the fragment energies, $E_f=\sum_\alpha E_\alpha$, is not in general equal to the total energy and the difference between the two quantities is defined as the partition energy, $E_p$.  The partition energy can be split into components,
\ben
\begin{aligned}
\label{e:Ep}
E_p[\{n_{\alpha}\}] = T_s^{\rm nad}[\{n_{\alpha}\}]+&V\ext^{\rm nad}[\{n_{\alpha}\}]\\+ E\H^{\rm nad} [\{n_{\alpha}\}] +&E\xc^{\rm nad}[\{n_{\alpha}\}]~~,\\
\end{aligned}
\een
where each non-additive functional is defined as the difference between the functional evaluated on the sum of fragment densities and the sum of the functional evaluated on each fragment density.  Thus, the non-additive functionals depend on whether the fragments are treated using ensembles or fractional-orbital occupation.  In the ENS case we have,
\ben
\begin{aligned}
\label{e:Fnad}
F^{\rm nad}[\{n_{i\alpha\sigma}\}] \equiv F[n_{f\uparrow},n_{f\downarrow}] - \sum_{i,\alpha}{f_{i\alpha}F_\alpha[n_{i\alpha\uparrow},n_{i\alpha\downarrow}]}.
\end{aligned}
\een
while in the FOO case we have,
\ben
\begin{aligned}
\label{e:Fnad_FOO}
F^{\rm nad}[\{n_{\alpha}\}] \equiv F[n_f] - \sum_{\alpha}{F_\alpha[n_\alpha]}.
\end{aligned}
\een  
Here is where we first encounter the NAKE functional within PDFT.  It is important to note that within standard Kohn-Sham approximations, the second, third and fourth terms of Eq. \ref{e:Ep} will be explicit functionals of the density.  However, in Kohn-Sham calculations the non-interacting kinetic energy is instead an implicit functional of the density.

Each fragment calculation proceeds via a KS system of non-interacting electrons in an effective potential, except with an additional potential, $v_p$, (the partition potential) which is the same for all fragments.
\ben
\{-\frac{1}{2}\nabla^2 + v^{\rm eff}_{\alpha}[n_{p_\alpha}](\br)+v_p(\br)\}\phi_{i,p_\alpha}(\br) = \epsilon_{i,\alpha}\phi_{i,p_\alpha}(\br)
\een
The partition potential is the functional derivative of the partition energy with respect to the total density and the chain rule is used to construct it via derivatives with respect to fragment densities,
\ben
v_p(\br)=\int\sum_\alpha\sum_{x=0,1}\frac{\delta E_{p}}{\delta n_{p_\alpha+x}(\br')}Q_{\alpha,x}(\br',\br)d^3\br',
\label{partition_potential}
\een
where the $Q$-functions are given by:
\ben
Q_{\alpha,x}(\br',\br)=\frac{\delta n_{p_\alpha+x}(\br')}{\delta n_f(\br)},
\een
and must satisfy the sum-rule\cite{MW2013}:
\ben
\sum_\alpha\sum_{x=0,1}Q_{\alpha,x}(\br',\br) = \delta(\br'-\br)
\label{sum_rule}
\een
We approximate the $Q$-functions via a local approximation:
\ben
Q^{\rm local}_{p_\alpha,x}(\br',\br) = \frac{n_{p_\alpha,x}(\br)}{n_f(\br)}\delta(\br-\br')
\label{eq:localQ}
\een

When we refer to the values of the kinetic component of the partition energy we refer to the NAKE functional of subsystem-DFT evaluated on the unique PDFT fragments.  However, when we consider the kinetic component of the partition potential this is not identical to the non-additive kinetic potentials of subsystem-DFT.  Instead they are related via the $Q$-functions,
\ben
v_{p,{\rm kin}}(\br)=\int\sum_\alpha\sum_{x=0,1}\frac{\delta T_{s}^{\rm nad}}{\delta n_{p_\alpha+x}(\br')}Q_{\alpha,x}(\br',\br)d^3\br'.
\label{vpkin}
\een
Because we use the local-$Q$ approximation these are in some sense approximate components of the exact partition potential.  However the local-$Q$ approximation obeys the sum rule of Eq. \ref{sum_rule} and therefore these components sum to the \emph{exact} partition potential.  Furthermore we can still make useful comparisons to approximate functionals using the same local-$Q$ approximation.

As mentioned earlier, for an approximate KS calculation, each component in Eq. \ref{e:Ep} is an explicit density functional of the fragment densities.  However, the exact NAKE functional is not.  The typical approach in fragment-based methods is to use an approximate density functional.  As we will see in section \ref{results}, approximations fail dramatically for covalently-bonded fragments.  Instead, we use numerical inversions to evaluate the exact NAKE functional for fragment densities.

Also important to PDFT is the preparation energy, $E_{\rm prep}$, which is the difference between the sum of fragment energies and the sum of isolated fragment energies, 
\ben
E_{\rm prep} = E_f-E_f^0
\een
The binding energy, the partition energy, and the preparation energy are all simply related via the equation:
\ben
E_{\rm bnd} = E_p+E_{\rm prep}
\een
The preparation energy can be broken into components, just like the partition energy.  These components will be displayed later in the results section in Table \ref{tab:energies}.

\section{Exact $T_s^{\rm nad}$ and its derivatives}
The exact $T_s$ defined via constrained search in Eq. \ref{eq:Ts} depends only implicitly on the density and depends explicitly on the Kohn-Sham orbitals.  In order to find the $T_s$ corresponding to a given density an inverse problem must be solved to find the orbitals minimizing the kinetic energy and yielding the density.  In order to find the functional derivative, $\frac{\delta T_s}{\delta n}$, we assume that the densities are solutions to the Euler equation,
\ben
\frac{\delta T_s}{\delta n(\br)} + v_s(\br) = \mu
\label{eq:euler}
\een
where $\mu$ is the chemical potential.  This equation can be solved for the functional derivative of $T_s$ so that if we have the chemical potential and effective potential corresponding to a given density we can calculate the derivative of the non-interacting kinetic energy with respect to that density.

This equation can be used directly for the fragments because the effective potential corresponding to each fragment is already available since it was used to calculate the fragment density.  However, the potential corresponding to the sum of the fragment densities is not immediately available and must be calculated via a numerical inversion the details of which will be described in the next section.  Once the inversion is complete, the functional derivative of the NAKE can be obtained by the following formula:
\ben
\frac{\delta T_s[\{n_\alpha\}]}{\delta n_\alpha(\br)} =  v_s[n_\alpha](\br) - v_s[n_f](\br) + \mu_f-\mu_\alpha
\label{eq:nakp}
\een
where $v_s[n]$ is the potential corresponding to the density $n$ and the $\mu_f$ and $\mu_\alpha$ are the chemical potentials which can be obtained from the HOMO eigenvalue corresponding to the numerical inversion and fragment $\alpha$ KS system respectively.

\section{Numerical Details}
Following the work of others\cite{Bec1982,MKK2009,KLS1996,LPS1983,GKG1997} our calculations are performed on a prolate-spheroidal real-space grid shown to give accurate results for all-electron calculations on diatomic molecules.  In this code, the Kohn-Sham equations are separated into a one-dimensional azimuthal problem and a two dimensional prolate spheroidal problem.  The azimuthal cooridinate is solved analytically, while the prolate spheroidal problem is solved numerically on a 5329 point grid.  We evaluate approximate exchange correlation and kinetic energy functionals using the Libxc library \cite{libxc}.

\subsection{Numerical Inversion}
  Our recipe for numerical inversion is essentially a direct search for orbitals solving the KS equations while simultaneously satisfying the necessary constraints.  The necessary constraints are that each orbital is correctly normalized, and that the sum of the squares of each orbital matches the target density.  We break these requirements into three groups, corresponding to satisfaction of the KS equations, the normilization constraint and the density constraint, and write down the corresponding residuals: 
\ben
\begin{aligned}
{\rm res}_{i,j}^{\rm KS} =& (-\frac{1}{2}\nabla^2\phi_j)_i + v_{{\rm eff},i}\phi_{j,i} - \epsilon_j\phi_{j,i}\\
{\rm res}_j^{\rm N} =& \sum_i \lvert \phi_{j,i} \rvert^2 - 1\\
{\rm res}_{i}^{\rm n} =&  \sum_j \lvert \phi_{j,i} \rvert^2 - n_i
\label{residuals}
\end{aligned}
\een  
where, the $i$ index runs over grid points and the $j$ index runs over orbitals.  $(-\frac{1}{2}\nabla^2\phi_j)_i$ is the finite difference approximation of the kinetic energry for orbital $j$ at grid point $i$, $v_{{\rm eff},i}$ is the effective potential at point $i$ and $n_i$ is the target density at point $i$.  For $N_p$ grid points and $N_m$ molecular orbitals, ${\rm res}_{i,j}^{\rm KS}$ ammounts to $N_p\times N_m$ equations ensuring satisfaction of the KS equations,  ${\rm res}_j^{\rm N}$ corresponds to $N_m$ equations ensuring orbital normalization, and ${\rm res}_{i}^{\rm n}$ corresponds to $N_p$ equations ensuring that the orbitals match the density constraint at each point on the grid.  These equations depend on orbitals and effective potential at each point on the grid as well as the eigenvalues.  We then combine Eq. \ref{residuals} to create a vector function which takes the orbitals, eigenvalues and effective potential as its argument.  Finding the root of this function ammounts to finding the effective potential corresponding to the given density.  The jacobian of this residual function is a sparse square matrix which can be used along with Newton-Raphson minimization to find the vector for which the residual is below some tolerance.
  We also make two modifications: first, we fix the HOMO eigenvalue to be equal to zero, which removes the ambiguity of shifting the effective potential by a constant and, second, we delete the normalization constraint on the HOMO and allow this constraint to be satisfied due to the overall density constraint. We found it helpful to enforce the normalization constraint at each iteration rather than to optimize it along with the other residuals.
\begin{figure}[htbp]
\includegraphics*[width=3.375in]{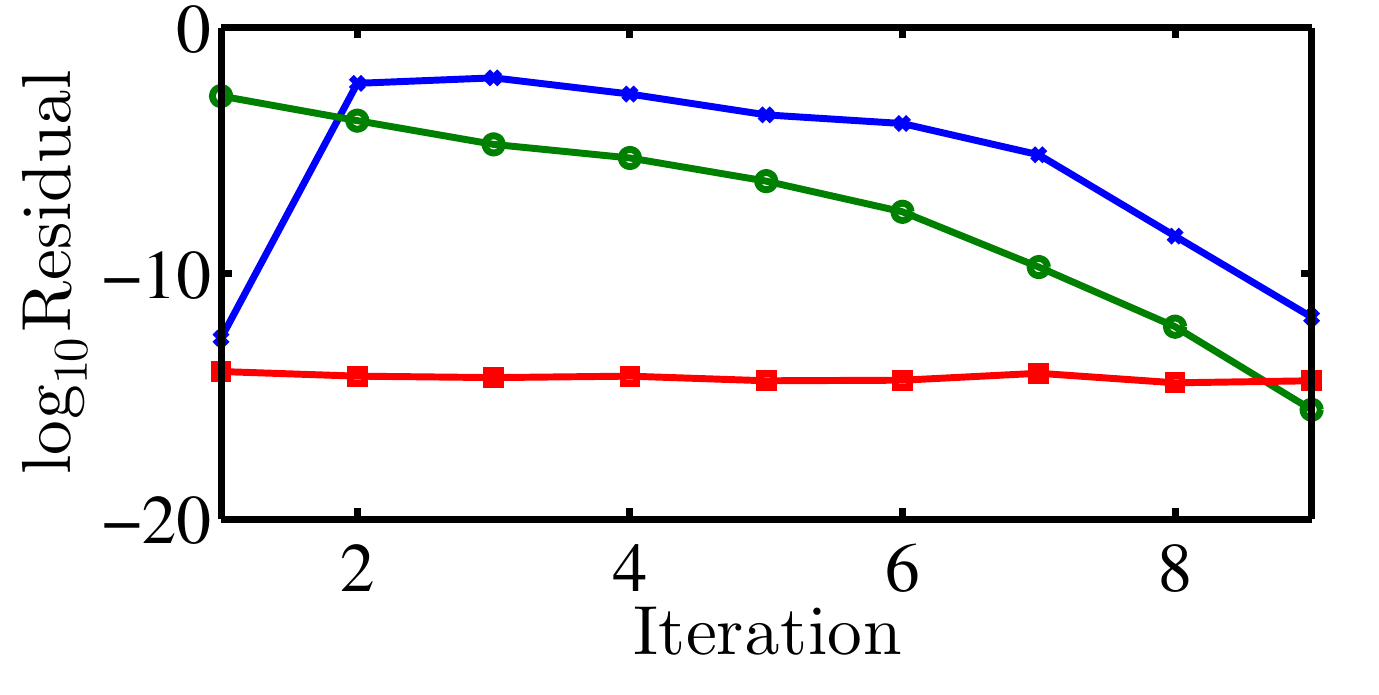}
\caption{The maximum absolute value of residuals of Eq. \ref{residuals} vs. iteration number during the inversion of isolated nitrogen atoms placed at the equilibrium distance of N$_2$.  Blue indicates how close the current orbitals are to solving the KS equation.  Green indicates indicates how close the density constraint is to being satisfied.  Red indicates how close the normalization constraint is to being satisfied.}
\label{fig:N2_inversion}
\end{figure}
We typically found that the inversions converged to a tolerance of less than $10^{-10}$ in around 10 iterations or less.  Figure \ref{fig:N2_inversion} shows the maximum residual vs. iteration for the inversion of N$_2$.  Our initial guess for the potential is simply the approximate Hartree exchange-correlation potential corresponding to the target density.  The orbitals corresponding to this potential form the initial guess for the potential.

These numerical inversions are used at each step in the PDFT SCF procedure to determine the functional derivatives of the NAKE with respect to the fragment densities used in the construction of the partition potential via Eq. \ref{partition_potential}. 

\subsection{PDFT SCF Convergence}

\begin{figure}[htbp]
\includegraphics*[width=3.375in]{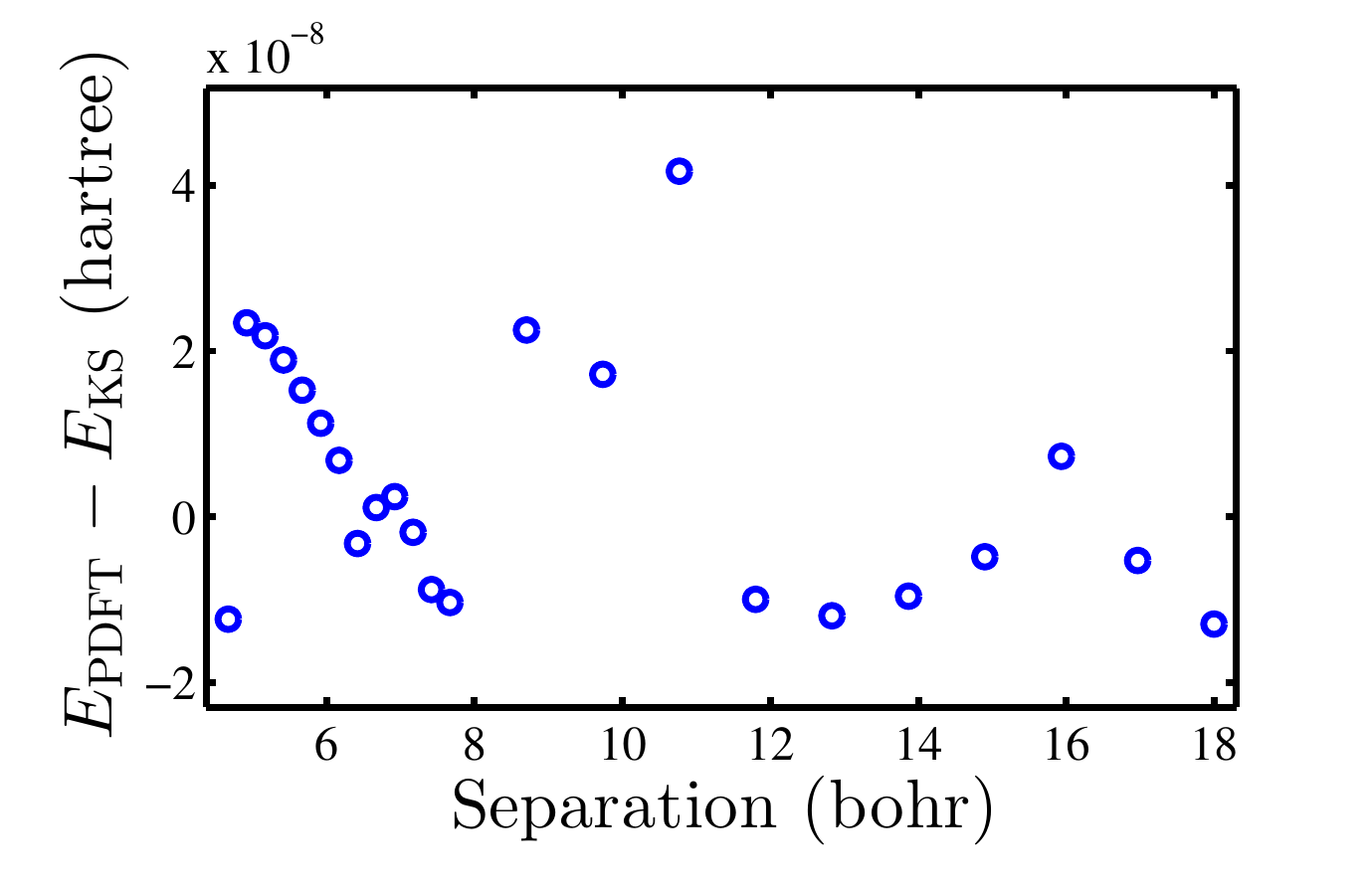}\\
\includegraphics*[width=3.375in]{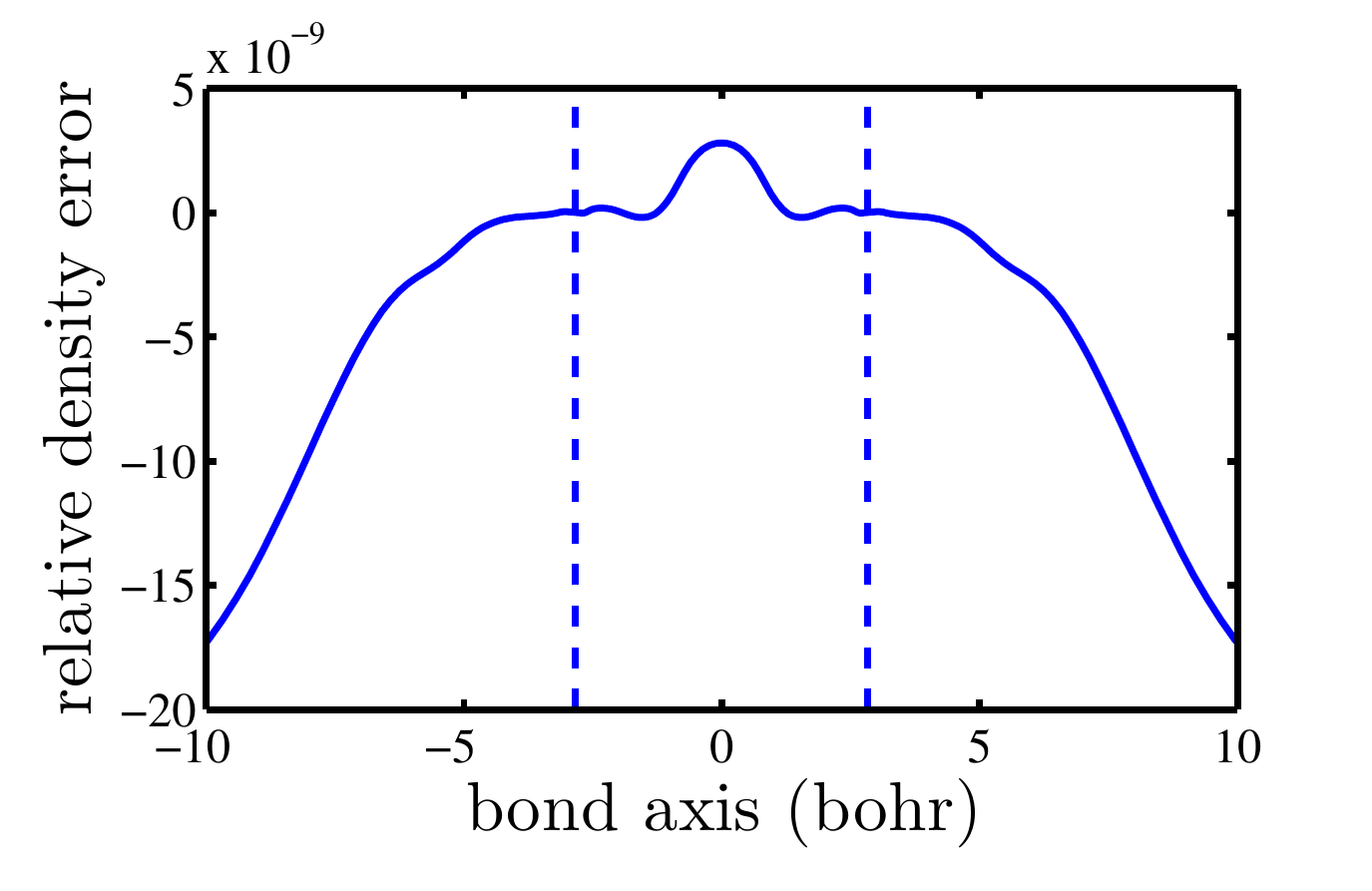}
\caption{Top: Energy error of PDFT calculation vs. Kohn-Sham calculation for Na$_2$ at various separations.  Bottom: Relative density error of PDFT calculation vs. Kohn-Sham calculation at equlibrium distance.}
\label{fig:Na2_error}
\end{figure}
\begin{figure}[htbp]
\includegraphics*[width=3.375in]{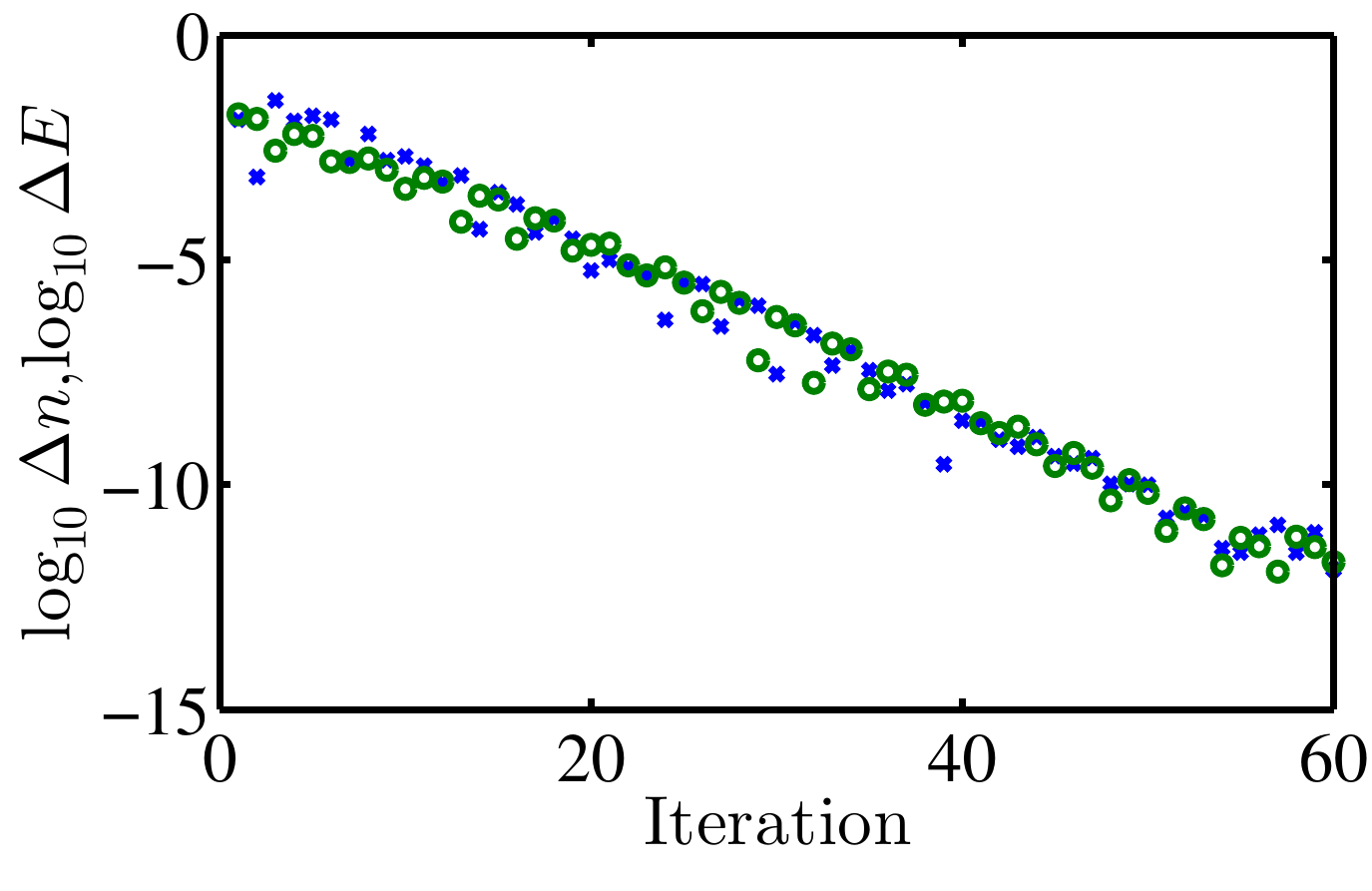}
\caption{$\Delta E =|E_{\rm PDFT}-E_{\rm KS}|$ (blue x's) and $\Delta n =\max|n_{\rm PDFT}-n_{\rm KS}|$ (green o's) vs. iteration number as the PDFT self-consistent procedure procedes.}
\label{fig:Li2_error}
\end{figure}
We run PDFT calculations for a wide range of separations on the H$_2$, Li$_2$, C$_2$, N$_2$, F$_2$ and Na$_2$ diatomics and use the LDA XC functionals throughout.

We first note that these calculations are capable of reproducing the corresponding KS calculations with remarkable accuracy.  Figure \ref{fig:Na2_error} displays the difference between KS-LDA and PDFT-LDA for energies and densities at the point where the SCF procedure was terminated for Na$_2$. It can be seen that energies match on the order of tens of nanohartree and the maximum relative density error is on the order of $10^{-8}$.  Figure \ref{fig:Li2_error} shows how the relative density and energy errors (as compared to KS-LDA) procede as a function of iteration number in the SCF procedure for Li$_2$.  Only a simple linear mixing scheme was used in the PDFT SCF procedure, but nonetheless Figure \ref{fig:Li2_error} shows good convergence.

It is worth noting that other studies presenting `accurate' non-additive kinetic potentials within codes using gaussian or slater type basis sets do not come close to this accuracy.  Goodpaster et al.\cite{GAMMI2010} report energy errors ranging from $0.09\%$ to $3.77\%$ for their exact DFT-in-DFT embedding method.  Fux et al. report the integral of the absolute value of the difference between KS densities and the densities obtained from their accurate frozen density embedding potentials to range between $0.0108$ and $0.1815$ (refered to as absolute error,$\Delta^{\rm abs} = \int |\rho_{\rm err} (\br)|d^3r$ in Table 1 of their paper\cite{FJN+2010}.  While these errors are clearly significantly better than those of approximate kinetic energy functionals and are therefore useful in evaluating errors in these functionals, our real-space code provides significantly more accurate results.  Our largest energy error at any separation was 1.9 microhartree for N$_2$ which corresponds to a $1.7\times 10^{-6}\%$ percent error and typical errors for our calculations were around 100 times smaller.  This is around $10^4$ times more accurate as compared to the results of Goodpaster et al.  Table \ref{tab:errors} shows the absolute error of our densities as compared to KS densities and we see that our largest absolute error is also about $10^4$ times more accurate than the smallest error reported by Fux et al.\cite{FJN+2010}  
\begin{table}[htbp]
\centering
\caption{Maximum error at any separation between PDFT-LDA and KS-LDA calculations performed on the same grid.  $\Delta E$ is the difference in total energy in hartree.  The density difference is measured by the integral of the absolute value of the density difference: $\Delta^{\rm abs} = \int |\rho_{\rm err} (\br)|d^3r$ in atomic units.}
\label{tab:errors}
\begin{tabular}{ccl}
\hline
 & $\max\ \Delta E $  & $\max\Delta {\rm abs}$ \\ \hline
H$_2$       & 3.3e-08  & 7.0e-08     \\
Li$_2$      & -1.7e-08  & 2.9e-08     \\
C$_2$       & -1.5e-06  & 7.5e-06     \\
N$_2$       & 1.9e-06  & 3.3e-06     \\
F$_2$       & -1.2e-08  & 1.4e-07     \\
Na$_2$      & 4.2e-08  & 4.4e-08    
\end{tabular}
\end{table}

This difference in accuracy is not due to signficicant difference in the method of inversion used by these groups but to the use of basis sets vs. real-space representation of densities and potentials.  Two of us\cite{NWW2011} previously  used similar inversion procedures performed with gaussian-type basis sets and achieved similar density errors to those of Fux et al.  In fact, Unsleber et al.\cite{UNJ2016} showed that while in the infinite basis-set limit it is possible for the sum of fragment densities to match total densities, with an incomplete basis set it may only be possible to find a near match between the fragment and total densities.  However, when the fragment and total densities are represented on real-space grids it is trivial to show that the total density can match the sum of fragment densities.

Another advantage that real-space grids have over gaussian basis sets for numerical inversions is that the corresponding potentials are uniquely represented on the same grids up to an additive constant.  For incomplete gaussian basis sets there may be many different spatial representations of the potential which yield the same density.  In this case, regularization procedures are used to unambiguously choose one potential \cite{Jac2011, FJN+2010,GRS2013}  With real-space grids this procedure is unnecessary and the representation of the potential on the grid can be used.

\section{Results}
\label{results}
\begin{table*}[htb]
\centering
\caption{Equilibrium bond lengths as well as bond lengths which minimize the exact ENS NAKE.  Also, LDA binding energy, partition energy, preparation energy, as well as components of the partition energy and preparation energy are given for the equilibrium distance.  The components are kinetic, classical coulomb (Hartree,electron-nuclear and nuclear-nuclear) and exchange-correlation).  All energy units are milihartree.}
\label{tab:energies}
\begin{tabular}{cccccccccccc}
\hline
    & $R_{\rm eq}$ & $R_{\min T_s^{\rm nad}}$ & $E_{\rm bind}$ & $E_p$     & $E_{\rm prep}$ & $T_s^{\rm nad}$ & $E_{\rm coul}^{\rm nad}$ & $E_{\rm xc}^{\rm nad}$ & $E_{\rm prep}^{\rm kin}$ & $E_{\rm prep}^{\rm coul}$ & $E_{\rm prep}^{\rm xc}$ \\ \hline
H$_2$  & 1.45 & 1.40      & -180.28& -225.58   & 45.31  & -152.06 & -71.76    & -1.77   & 302.32    & -169.67    & -87.34   \\
Li$_2$ & 5.18 & 7.53      & -37.54 & -49.16  & 11.62  & 3.22    & -33.08    & -19.31  & 17.37     & 6.67       & -12.41   \\
C$_2$  & 2.34 & 3.15      & -267.21& -448.68 & 181.47 & 19.90   & -353.66   & -114.92 & 233.61    & 27.70      & -79.84   \\
N$_2$  & 2.07 & 3.17      & -427.00& -549.73 & 122.73 & 180.20  & -622.05   & -107.89 & 189.42    & 27.91      & -94.60   \\
F$_2$  & 2.62 & 3.41      & -124.91& -157.57 & 32.66  & 0.86    & -100.93   & -57.50  & 101.32    & -24.84     & -43.82   \\
Na$_2$ & 5.69 & 8.62      & -33.15 & -40.65 & 7.50   & 17.01   & -34.65    & -23.02  & -0.18     & 6.44       & 1.24        
\end{tabular}
\end{table*}
Table \ref{tab:energies} compares values of $E_p$ and $E_{\rm prep}$ components.  These components provide a bond-decomposition analysis similar in spirit to the bond-decomposition scheme of Ruedenberg \cite{Rue1962,SIR2014}.  The preparation energy is similar to the intra-atomic deformation energy as it indicates the ammount of energy required to deform isolated fragments into fragments within the molecule.  It is also similarly always positive.  The partition energy is similar to the inter-atomic energy lowering.  The sum of the intra-atomic deformation energy and the inter-atomic energy lowering is equal to the binding energy just as the partition energy plus the preparation energy also equals the binding energy.  The differences center around the fact that our fragment decomposition is based on the electronic density rather than orbitals, as is the case for the quasiatoms of Schmidt et al. \cite{SIR2014}.  Furthermore, our results are based on Kohn-Sham calculations and therefore we have access to the non-interacting kinetic energy rather than the total kinetic energy.  The non-interacting kinetic energy is governed by the KS virial theorem and not the more general virial theorem.  
Nevertheless, the components of the preparation energy do behave in a similar fashion to the components of the intra-atomic deformation.  Except for Na$_2$, the kinetic and hartree components are always positive indicating that during bonding the fragments contract as discussed in reference \citenum{SIR2014}.

Of course, all these calculations are performed at the LDA level of theory, so we take these analysis with a grain of salt.  The primary goal of this study is to look at the non-additive kinetic energy and corresponding potentials and how well they reproduce a DFT calculation for a given XC functional regardless of its intrinsic accuracy.

\subsection{Non-additive kinetic energies}

\begin{figure}[htbp]
\includegraphics*[width=3.375in]{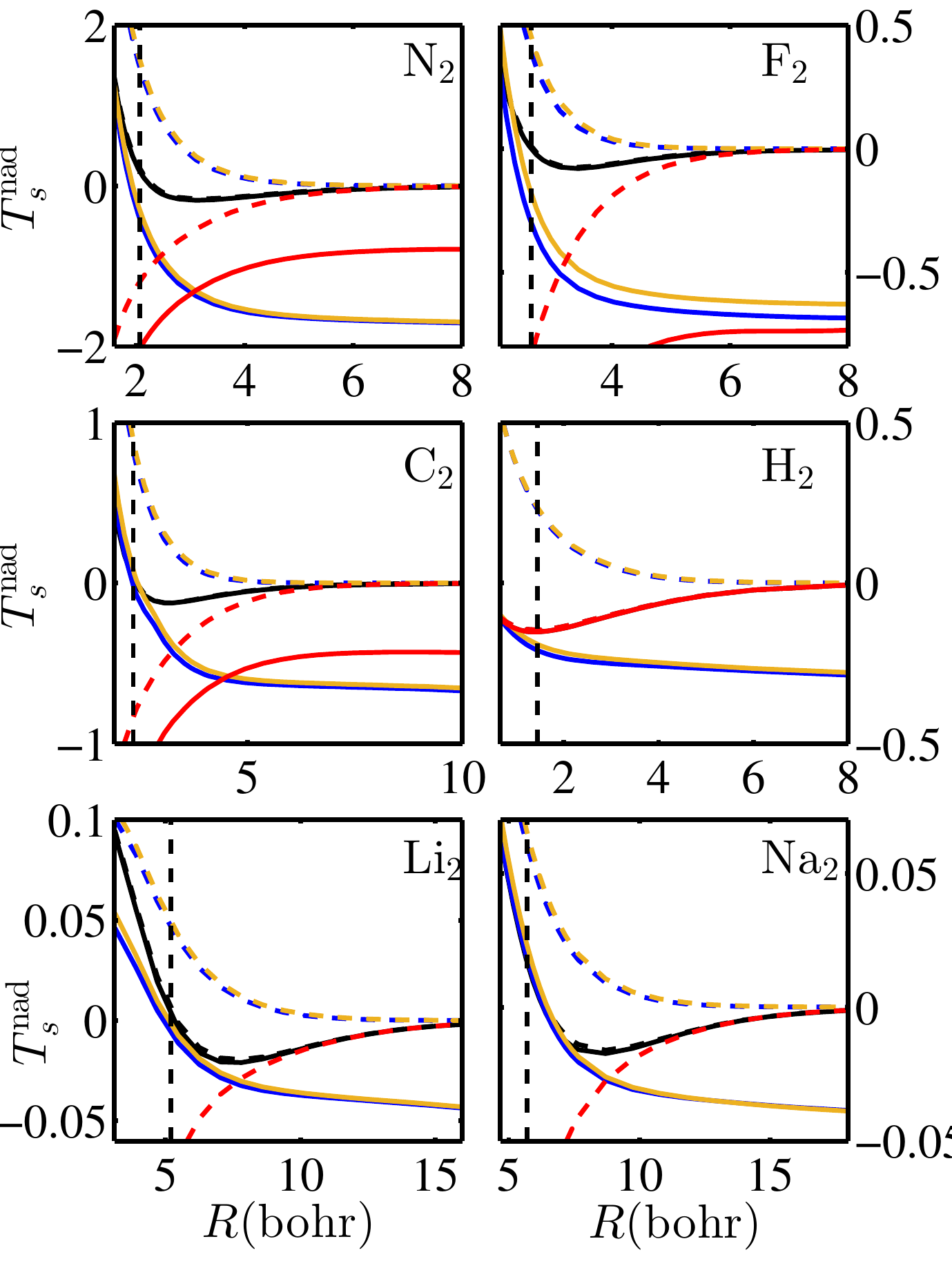}
\caption{Exact non-additive kinetic energy for ensemble (black solid) fragments vs. fractionally occupied orbital fragments (black dashed).  Note that these two curves are nearly indistinguishable.  Several approximate kinetic energy functionals are also shown: LC94 in blue, Thomas-Fermi in orange), and von Weiz{\"a}cker in red.  Each of these is plotted for for ensemble fragments (solid lines) and fractionally occupied orbital fragments (dashed lines).  The LDA XC functional is used for all calculations. }
\label{fig:ENSvsFOO}
\end{figure}

Figure \ref{fig:ENSvsFOO} displays the non-additive kinetic energies for all the systems studied.  The equilibrium separation is also plotted as a vertical line.  In all cases, the non-additive kinetic energy displays a minimum separation.  The separation where this minimum occurrs is displayed in Table \ref{tab:energies}.  Except for H$_2$, this minimum occurs significantly outside the equilibrium separation (from about 1.3 to 1.5 times the equilibrium distance). Thus, at equilibrium distances the non-additive kinetic energy constitutes a repulsive force, while at very large separations the non-additive kinetic energy acts in an atractive manner.  In contrast, for H$_2$, the minimum in the non-additive kinetic energy is at roughly the same distance as the equilibrium.

\subsection{Ensemble vs. Fractional Orbital Occupation}
Figure \ref{fig:ENSvsFOO} reveals that the exact non-additive kinetic energy using the ensemble-based fragments is very close to the exact non-additive kinetic energy when using the fractionally-occupied orbitals.  Indeed, the two curves are nearly indistinguishable.  However, for the approximate functionals there is a very large difference between the two methods.  These differences are due to the incorrect treatment of fractional spins by approximate density functionals.  

Except for in the case of H$_2$, where the von Weiz{\"a}cker functional is actually exact, it can be seen that the approximate functionals evaulated on both the ensemble and fractionally-occupied orbital fragments are quite innacurate.  Neither ensemble nor fractional orbital occupation fragments have a  minimum as displayed by the exact non-additive kinetic energy.  Instead, all the approximate functionals yield values which decrease monotonically with separation.  

However, it is interesting to note that close to equilibrium, Thomas-Fermi and the LC94 functionals evaluated on the ensemble densities significantly outperform the same functionals evaluated on the fractionally occupied orbital fragments.  Table \ref{tab:ENSvsFOO} compares results for ENS and FOO treatments using both Thomas-Fermi and the LC94 functional.  The results show that both approximate functionals evaluated using the ENS treatment are more accurate at equilbrium separations.  From Figure \ref{fig:ENSvsFOO} it appears that for most of the systems TF and LC94 using ENS becomes more accurate at even smaller separations.  At large separations the opposite is true with the FOO treatment of approximate functionals becoming more accurate.  The pure von Weiz{\"a}cker functional evaluated on FOO fragments performs especially well at large separations and for FOO fragments it is the only approximate functional that gives the correct sign.    

\begin{table}[htb]
\centering
\caption{Comparison of exact vs. approximate NAKE functionals using the ENS vs. FOO treatment of fragments at the equilibrium distance for each molecules.  Units are milihartree.}
\label{tab:ENSvsFOO}
\begin{tabular}{ccccccc}
      & \multicolumn{2}{c}{Exact} &  \multicolumn{2}{c}{Thomas-Fermi} & \multicolumn{2}{c}{LC94} \\ \hline
      & ENS         & FOO             & ENS             & FOO            & ENS         & FOO        \\ \hline
H$_2$  & -152.06   & -145.43          & -190.92       & 234.75       & -209.80   & 227.98   \\
Li$_2$ & 3.22      & 6.09              & -1.91         & 49.44        & -5.89     & 46.59    \\
C$_2$  & 19.90     & 42.80          & 77.09         & 884.70       & -20.72    & 822.80   \\
N$_2$  & 180.20    & 218.59         & -299.91       & 1572.57      & -408.17   & 1480.73  \\
F$_2$  & 0.86      & 10.91          & -178.90       & 455.59       & -298.34   & 391.09   \\
Na$_2$ & 17.01     & 19.82          & 23.21         & 65.47        & 18.32     & 60.58   
\end{tabular}
\end{table}

\subsection{Kinetic component of the partition potential}

\subsubsection{Na$_2$ and Li$_2$}

\begin{figure*}[htbp]
\includegraphics*[width=6.38in]{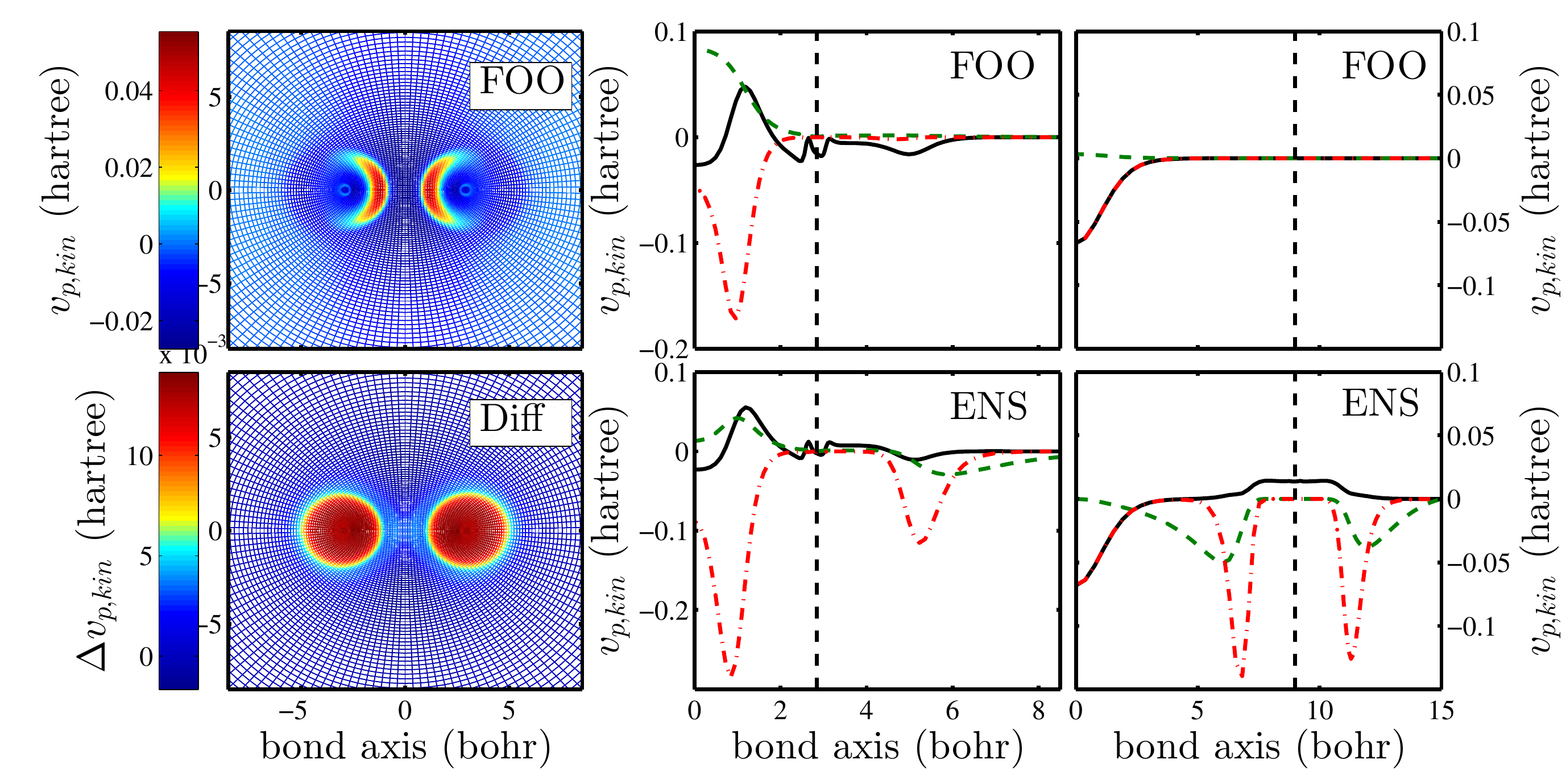}
\caption{Kinetic portion of the partition potential for Na$_2$.  The left column shows plots on the bonding plane, on the prolate spheroidal mesh, with the top plot showing the kinetic component using FOO and the bottom plot showing the difference between the kinetic component evaulated using FOO and ENS.  The right two columns plot exact kinetic components (black solid) vs. Thomas-Fermi (green dashed) and von Weiz{\"a}cker (red dot-dashed) along the bond axis.  Because there is mirror symmetry only the right side is plotted.  The location of the nuclei is indicated by a vertical black dashed line.  The middle and left columns show results for the equilibrium separation ($R = 5.69$ bohr) and the right column shows a stretched configuration ($R = 18.0$ bohr). The top row uses FOO and the bottom row uses ENS.}
\label{fig:vp_kin_Na2}
\end{figure*}
\begin{figure*}
\includegraphics*[width=6.38in]{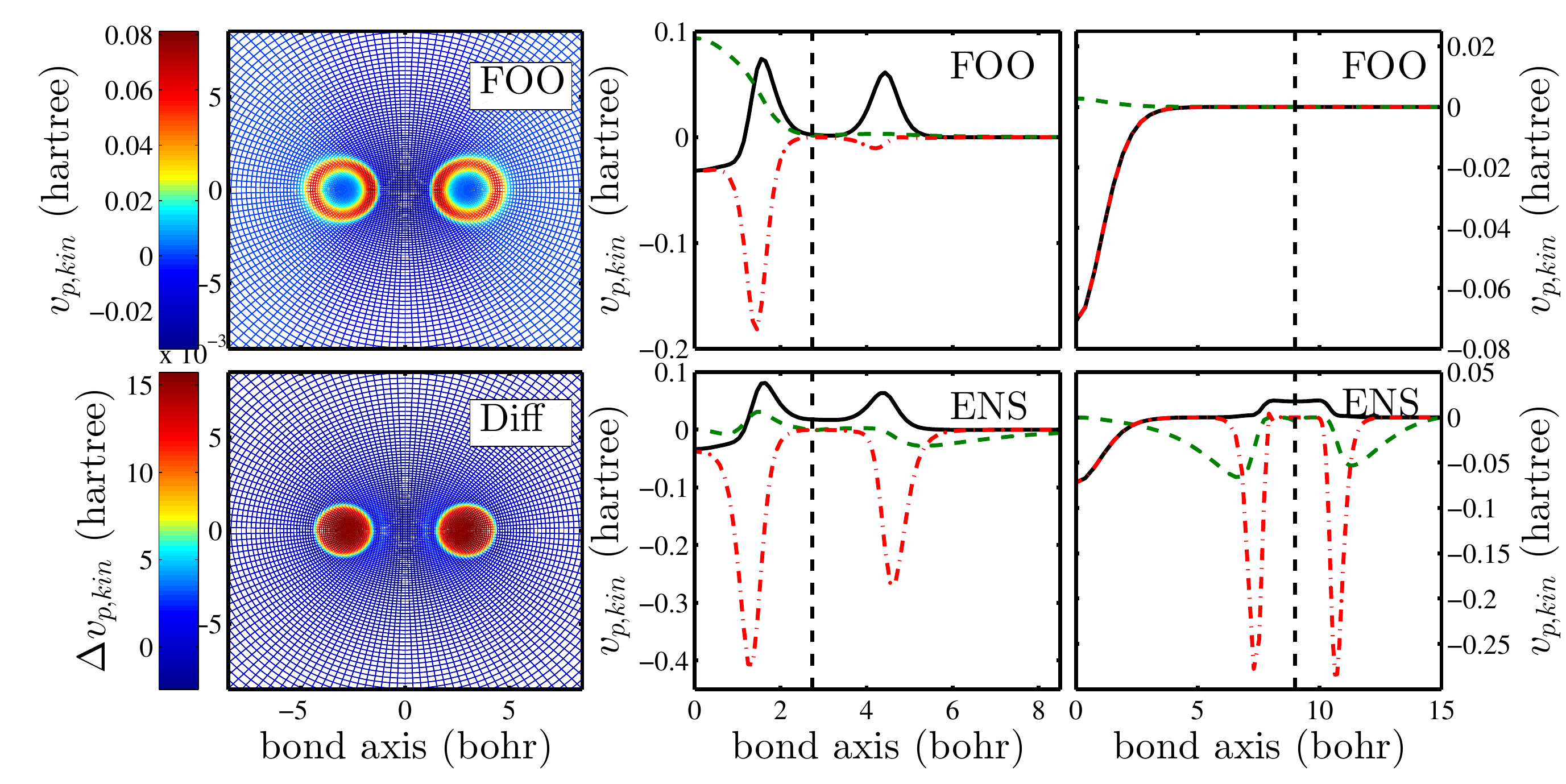}
\caption{Same as figure \ref{fig:vp_kin_Na2} with results for Li$_2$.  Equilibrium separation is at 5.18 bohr (middle and left columns) and stretched configuration is at 18.0 bohr (right column).}
\label{fig:vp_kin_Li2}
\end{figure*}
\begin{figure*}
\includegraphics*[width=6.38in]{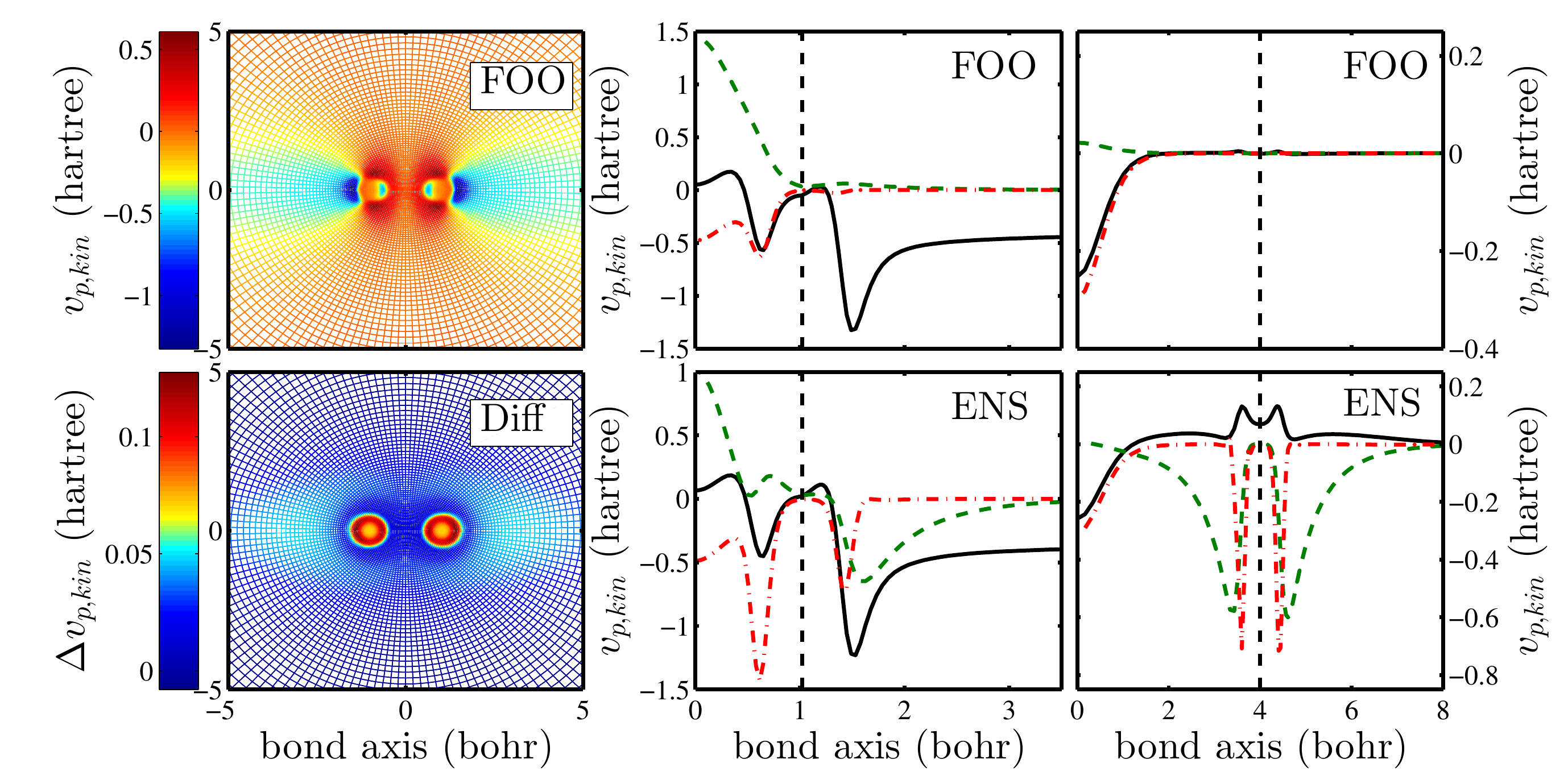}
\caption{Same as figure \ref{fig:vp_kin_Na2} with results for N$_2$.  Equilibrium separation is at 2.07 bohr (middle and left columns) and stretched configuration is at 5.72 bohr (right column).}
\label{fig:vp_kin_N2}
\end{figure*}
\begin{figure*}[htb]
\includegraphics*[width=6.38in]{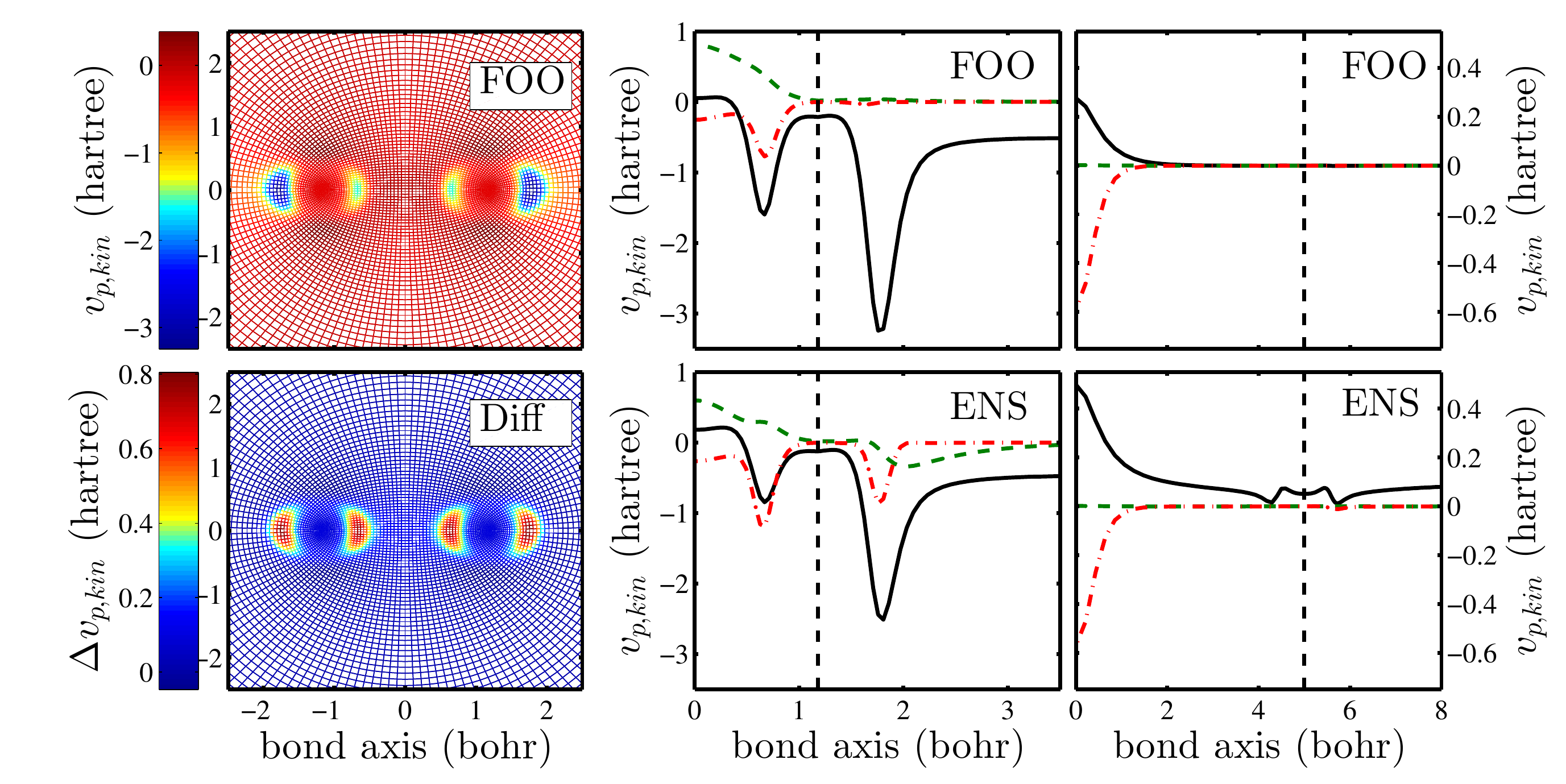}
\caption{Same as figure \ref{fig:vp_kin_Na2} with results for C$_2$.  Equilibrium separation is at 2.34 bohr (middle and left columns) and stretched configuration is at 10.0 bohr (right column).}
\label{fig:vp_kin_C2}
\end{figure*}
\begin{figure*}[htb]
\includegraphics*[width=6.38in]{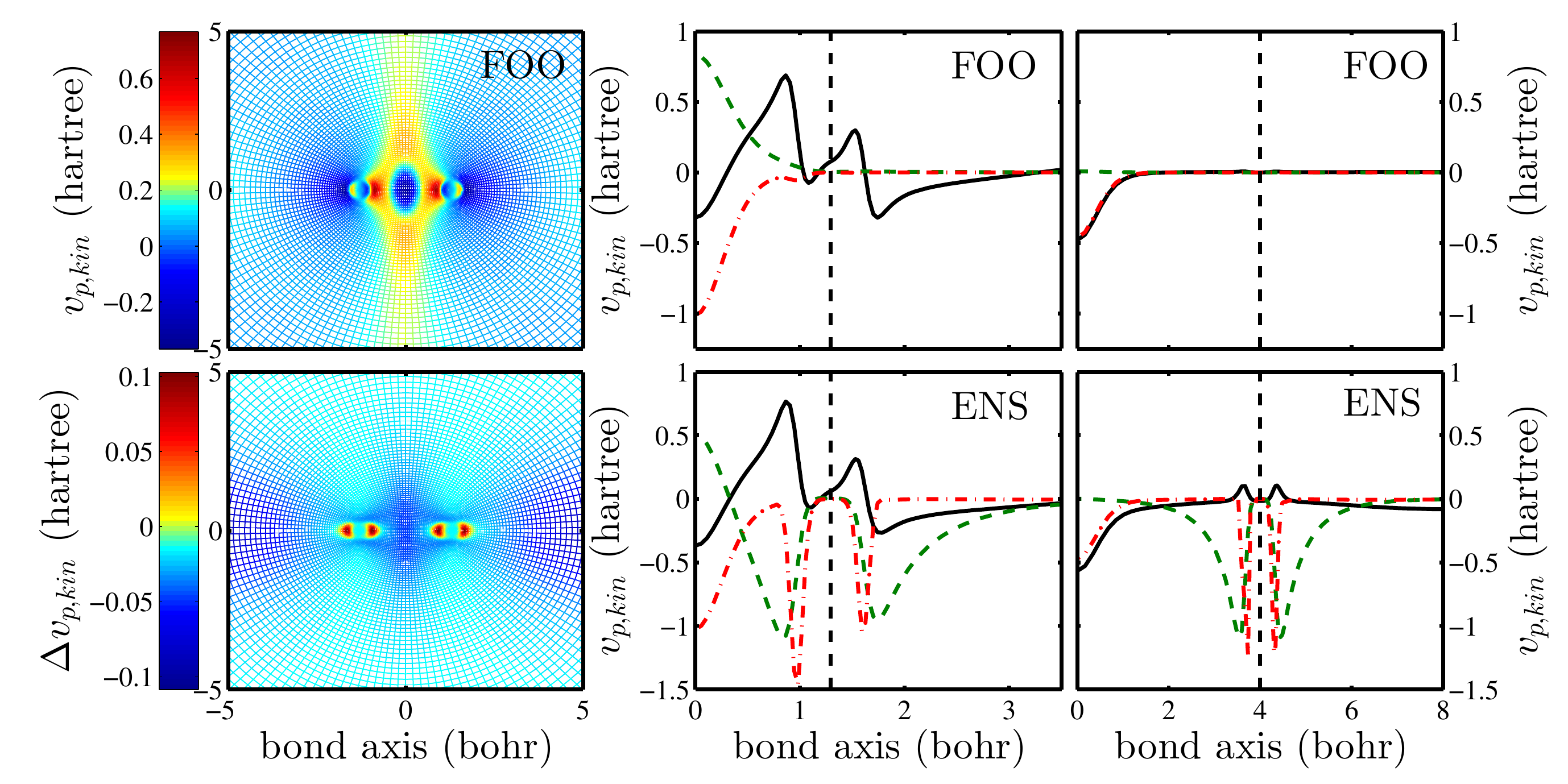}
\caption{Same as figure \ref{fig:vp_kin_Na2} with results for F$_2$.  Equilibrium separation is at 2.62 bohr (middle and left columns) and stretched configuration is at 8.0 bohr (right column).}
\label{fig:vp_kin_F2}
\end{figure*}
In both Na$_2$ and Li$_2$, the difference between the kinetic component of the partition potential for FOO and ENS fragment treatment is quite small and is plotted in the lower left of Figures \ref{fig:vp_kin_Na2} and \ref{fig:vp_kin_Li2}.  Their difference appears to form small, relatively flat plateaus surrounding each atom, approximately over the region of space where the core electron density dominates over the valence density.   This is also clear in the stretched case (right-most column), where the exact ENS $v_{p,{\rm kin}}$ has a plateau feature near the fragment and the exact FOO $v_{p,{\rm kin}}$ has none.    This feature has a straightforward explanation.  The correct dissociation products for Na$_2$ are open-shell sodium atoms similar to the components of the ensemble.  However, LDA XC functional suffers from static-correlation error and so as the Na$_2$ and Li$_2$ bonds stretch in an LDA calculation, the fragment densities are incorrect.  These incorrect fragment densities resemble the FOO atoms.  Thus, the kinetic component of $v_p$ in the FOO case is flat while in the ENS case it must compensate to arrange the ensemble components to match the incorrect stretched LDA density.  

As was the case for the non-additive kinetic energy, the difference between ENS and FOO representations is much more pronounced when considering the Thomas-Fermi and von Weiz{\"a}cker approximations due to the incorrect treatment of fractional spins.  The kinetic components of the partition potential corresponding to these approximations are also plotted in Figure \ref{fig:vp_kin_Na2}.  Both von Weiz{\"a}cker and Thomas-Fermi evaluated on FOO fragments appear simpler and only have significant features in the region in between the nuclei. The approximations evaluated on the ENS fragments show significant features near the fragment even in the stretched case.  These features in the ENS von Weiz{\"a}cker often appear in the same location as features in the exact $v_{p,{\rm kin}}$, but they are often significantly different in magnitude and even sign.

However, it is quite interesting to note the region in the stretched case where the von Weiz{\"a}cker approximation is nearly perferct for both Na$_2$ and Li$_2$.  In the case of the FOO fragments it is nearly perfect everywhere.  Na$_2$ and Li$_2$ have a single valence electron and it appears that the von Weiz{\"a}cker performs exceptionally well in regions where both fragments are dominated by their valence density.

\subsubsection{N$_2$ C$_2$ and F$_2$}

Figures \ref{fig:vp_kin_N2},\ref{fig:vp_kin_C2} and \ref{fig:vp_kin_F2} compare the ENS vs. FOO $v_{p,{\rm kin}}$ for N$_2$, C$_2$, and F$_2$ respectively.  For these stronger bonds the corresponding kinetic components of the partition potential are roughly 10 to 30 times stronger than for the alkali dimers, as are the corresponding differences between the exact $v_{p,{\rm kin}}$ for ENS and FOO.  Nevertheless the comparison between $v_{p,{\rm kin}}$ in the stretched FOO case vs. stretched ENS is quite similar, with stretched FOO $v_{p,{\rm kin}}$ remaining quite flat while stretched ENS retaining significant features.  The same argument about static correlation error in LDA applies to these cases to explain the ENS vs. FOO differences. 

For these dimers, the exact kinetic component of the partition potential has significantly more features in the valence region.  N$_2$ has a significant trough that extends along the bond axis in the regions outside the bond area.  On the other hand, F$_2$ has a ridge in the perpendicular direction along the bonding midplane.  In addition to these features there are also several peak and valley features in the region around the density transitions between core-dominated to valence-dominated.  

Except for in C$_2$, the von Weiz{\"a}cker functional still does a good job around the bonding midpoint in the stretched configurations, although not as well as for the alkali dimers.  For C$_2$, the exact kinetic component of the partition potential has a peak at this point while all other systems we considered had a well and were described reasonably well by the von Weiz{\"a}cker approximation.

\section{Conclusions}

As is well known, NAKE functional approximations perform very poorly for strongly overlapping fragments such as covalent bonds.  Here, we have provided exact and uniquely defined reference data for the behavior of the NAKE for several covalently-bonded diatomics, with the goal of guiding the development of new approximations for the NAKE functional.  

Our real-space numerical inversions give more accurate and unambiguous results than inversions based around basis sets used in previous studies of exact non-additive kinetic potentials\cite{GAMMI2010,FJN+2010}.  Also, as compared to analytical studies of non-additive kinetic potentials for four electron systems\cite{DSW2012} our paper addresses significantly larger systems, while retaining high accuracy. 

We found that the exact NAKE evaluated on PDFT fragments has a bonding type of behavior where it is attractive at large separations and repulsive at close separations near equilibrium.  We also compared two methods for handling fractional spins: the fractional orbital occupation method (FOO) and the PDFT ensemble method (ENS) and ENS methods for both exact and approximate NAKE functionals.  We found that FOO appears more accurate at larger separation while ENS is more accurate near equilibrium.

In addition to the NAKE, we also looked at the kinetic component of the partition potentials which are directly related to the non-additive kinetic potentials used in subsystem-DFT.  We found significant regions where von Weiz{\"a}cker gives nearly exact results for the kinetic component of the partition potential.

{\em Acknowledgments:} 
This work was supported by the Office of Basic Energy Sciences, U.S. Department of Energy, under grant No.DE-FG02-10ER16196.  A.W. also acknowledges support from the Camille Dreyfus Teacher-Scholar Awards Program.

\bibliography{NJW16}

\end{document}